# Theory of Remaining Exceptional Points from Nongeneric Splitting in Non-Hermitian Systems

Teng Yin,[1] Hao Zhang,[2] and Wenxiu Li[3]

[1]*School of Electronic Information Engineering, Beihang University, Beijing 100191, China* [2]*SchooL of Space and Earth Sciences, Beihang University, Beijing 100191, China*
[3]*School of Physics, Beihang University, Beijing 100191, China*

In non-Hermitian physics, high-order exceptional points(HOEPs) with eigenvalues and eigenvectors coalesce are known for their enhanced sensitivity to perturbations. Typically, they exhibit eigenvalue splitting that scales as $\varepsilon^{1/n}$, which is referred to as the generic response. However, under certain conditions, a nongeneric response of HOEPs occurs where the splitting follows a lower order $\varepsilon^{\frac{1}{m}}(m<n)$. A nongeneric response of HOEPs with a lower order splitting lead to the remaining EPs. While the presence of these remaining EPs is acknowledged, a thorough elucidation of their fundamental properties has yet to be achieved. In this work, we demonstrate those unsplit eigenvalue points must constitute remaining EPs in a perturbed n-orders HOEPs system. Combining graph theory and topological analysis, the number and splitting order of the remaining EPs is studied. This framework not only resolves a fundamental challenge in HOEPs but also paves the way for exploiting remaining EPs in applications such as anisotropic sensing and the design of Dirac exceptional points.

## I. Introduction

Non-Hermitian physics has come to the fore as a crucial framework for the exploration of open systems, spurring exciting progress in the fields of microwave billiards[1], acoustic resonances[2], parity-time(PT) or anti-PT symmetric coupled system[3,4], optical microcavities[5] and so on. A unique feature of non-hermitian systems is the excistence of singularities known as exceptional points(EPs). There are many novel physical phenomena are related to EPs, such as coherent perfect absorption[6], non-reciprocal devices[7,8], optical nonlinearity[9], novel optical amplifiers[10] and EP-based sensing[11-18]. The key characteristic—EPs—demonstrates groundbreaking potential in the field of photonic sensing, where the simultaneous degeneracy of eigenstates and eigenvalues endows EP systems with a unique nonlinear response to external perturbations[19]. In an nth-order $EP_n$ system, where n eigenvalues and their corresponding eigenvectors coalesce, the application of a perturbation of strength $\varepsilon$ leads to a splitting of the eigenenergies that scales as $\varepsilon^{1/n}$. So a small perturbation $\varepsilon \ll 1$ and higher-order n in EP-based sensors will cause a larger response than linear splitting of conventional sensing.

While second-order EPs have been extensively explored in sensing applications, HOEPs remain incompletely understood due to their complex configurations and properties. Theoretically, A $HOEP_n$ system influenced by a perturbation exhibits eigenvalue splitting scaling as $\epsilon^{1/n}$, it is termed generic. However, such a generic response is not universal. A particularly illustrative example arises in an sixth-order EP of PT symmetric electronic circuit[20], exhibiting eigenvalue splitting proportional to $\epsilon^{1/4}$. This behavior is classified as nongeneric, a phenomenon often linked to the occurrence of incomplete eigenvalue splitting, leaving behind a set of remaining exceptional points(REPs). While the existence of such remaining EPs is recognized, a comprehensive understanding of their fundamental properties remains elusive.

In this work, we first employ mathematical methods to rigorously prove that the unsplit eigenvalues in a perturbed HOEP system must indeed form such remaining EPs. A graph-theoretical approach is employed to accurately characterize the numbers of remaining EPs, and their splitting order under further perturbations is analyzed by a novel topological trajectory theory.

## II. Conditions for the occurrence of remaining EPs

Consider a general n × n Hamiltonian $H = H_0 + \varepsilon H_1$, where $H_0$ hosts an $n$th-order exceptional point($EPn$) with a degenerate eigenvalue assumed to be zero(the conversion can be performed via $(H_0 - \lambda I)$) and $\varepsilon H_1$ is a perturbation matrix. As established in the **APPENDIX A**, the rank of the perturbed Hamiltonian is constrained by the inequality:

$$n - 1 \leq rank(H) \leq n \tag{1}$$

This constraint dictates the splitting behavior of the eigenvalues.

(i) **Full Rank**($rank(H) = n$): it is evident that all eigenvalues have undergone splitting at this condition. Otherwise, there would be zero eigenvalue, but this is impossible, as the existence of a zero eigenvalue would result in $rank(H) \neq n$.

(ii) **Rank Deficiency**( $rank(H) = n - 1$ ): there must exist at least one zero eigenvalue;otherwise, the matrix should be full-rank. It is easy to know that the number of linearly independent eigenvectors corresponding to the zero eigenvalue is $n - rank(H) = 1$, from which it can be concluded that the points without split eigenvalues in the perturbed Hamiltonian $H$ must constitute remaining EPs.

Since nongeneric response typically does not cause all eigenvalues to split, it can be inferred that the occurrence of remaining EPs is invariably accompanied by it. In summary, a rank deficiency of the perturbed Hamiltonian is both necessary and sufficient for the emergence of remaining EPs. Regarding the number and the splitting rules of the remaining EPs after perturbation, the subsequent section provides detailed solutions.

## III. Determine the number of remaining EPs

Having established that eigenvalue incomplete splitting leaves behind remaining EPs, a fundamental question arises: how many of such degeneracies persist? Although diverse methods exist for constructing HOEPs[21-24], any matrix hosting HOEPs can be transformed into the Jordan canonical form via a similarity transformation. We therefore begin our analysis with a Jordan block $H^*$, first addressing the case of a single perturbation before generalizing to scenarios involving multiple perturbations.

### A. Evaluation under a Single Perturbation

Prior to delving into the core theory, it is imperative to introduce essential foundational knowledge. If the eigenvalues of $H$ exhibit a splitting with a maximum multiplicity of $\varepsilon^{1/m}$, while the remaining $(n - m)$ eigenvalues do not split, the characteristic equation of the system must then take the form:

$$\lambda^{n-m}\left(a_0 \lambda^m + a_1 \lambda^{m-1} + \cdots\cdots + a_{m-1}\lambda + a_m\right) = 0 \tag{2}$$

where $a_m$ is a quantity related to the perturbation strength $\varepsilon$. This indicates that the formation of the highest-order splitting $\varepsilon^{1/m}$ requires the involvement of $m$ eigenvalues, leaving $(n - m)$ remaining EPs.

A particularly tractable scenario arises when the single perturbation is applied to a specific element of the Jordan block. The reference[25] indicates that if the single perturbation is applied to the $s$-diagonal entries $(H^*)_{j+s,j}$, where $s \in \{0,1,\ldots,n-1\}$, $H^*$ is Jordan Block, then the eigenvalues will split to form $\lambda \sim \varepsilon^{1/m}$, where$(m = s + 1)$. Consequently, the number of remaining EPs is $n - (s + 1)$. The $H_a$-matrix from

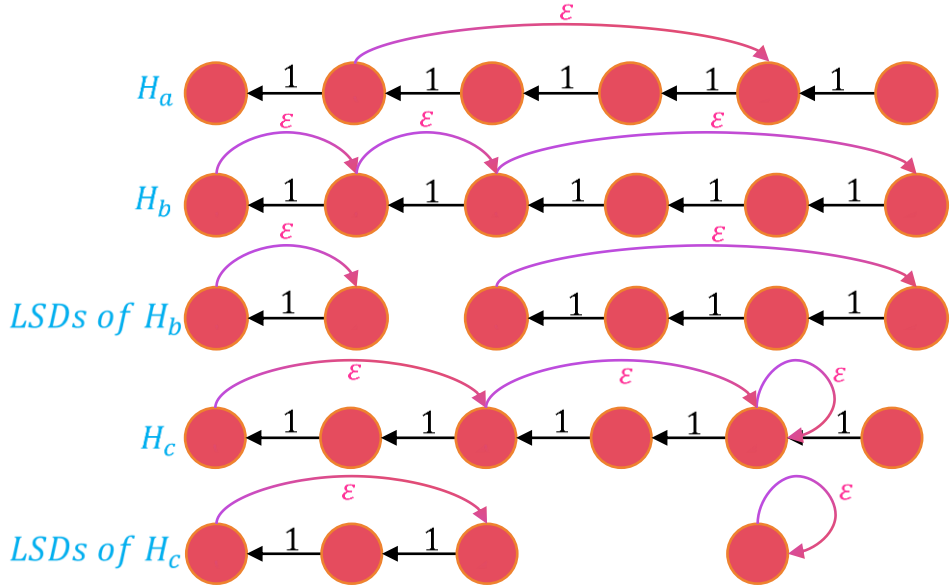

FIG. 1 The graph theory of matrices $H_a$, $H_b$ and $H_c$, respectively. In the graph plotted from the matrix, the colored edges represent the disturbance terms $\varepsilon$, while the black edges denote the constant terms.

Eq. (3) is employed as a case study to illustrate the proposed method.

$$H_a = \begin{pmatrix} 0 & 1 & 0 & 0 & 0 & 0 \\ 0 & 0 & 1 & 0 & 0 & 0 \\ 0 & 0 & 0 & 1 & 0 & 0 \\ 0 & 0 & 0 & 0 & 1 & 0 \\ 0 & \varepsilon & 0 & 0 & 0 & 1 \\ 0 & 0 & 0 & 0 & 0 & 0 \end{pmatrix} \quad (3)$$

A single perturbation at the fifth row and second column of the matrix is denoted by $\varepsilon_{5,2}$. A detailed graphical representation of matrix A is provided in FIG. 1. It can be concluded that $\lambda \sim \varepsilon^{1/4}$, the number of remaining EPs of $H_a$ is 2. This constitutes a rapid evaluation approach; however, its applicability is confined to single perturbation instances, while assessment of multiple perturbations remains challenging.

**B. Evaluation under Multiple Perturbations**

To address the determination of the quantity of remaining EPs under multiple disturbances, it is necessary to introduce the concept of a linear subdigraph(LSD)[26]. Given a graph G corresponding to a matrix, LSD is obtained by deleting any number of edges and vertices while ensuring that the retained vertices have exactly one in-degree and one out-degree(a self-loop counts as one in-degree and one out-degree). Note that LSDs are not necessarily unique. The standard practice for determining the number of remaining EPs is to use the following formula[26]:

$$b_{\mu,\beta} = \sum_{L \in \mathcal{L}_{\mu,\beta}(H)} (-1)^{c(L)} w(L) \quad (4)$$

where $L_{\mu,\beta}(H)$ denotes the set of LSDs with $\beta$ vertices and $\mu$ perturbation edges, $c(L)$ is the number of disconnected components, and $w(L)$ is the product of its edge weights.

Prioritize finding an LSD that maximizes the value of $\beta$ as much as possible. On this basis, further seek to minimize the perturbed edge $\mu$. If the calculated value of $b_{\mu,\beta} \neq 0$, it can be concluded that the number of remaining EPs is $(n - \beta)$. A representative matrix $H_b$ from Equation(5) will be used for specific illustration.

$$H_b = \begin{pmatrix} 0 & 1 & 0 & 0 & 0 & 0 \\ \varepsilon & 0 & 1 & 0 & 0 & 0 \\ 0 & \varepsilon & 0 & 1 & 0 & 0 \\ 0 & 0 & 0 & 0 & 1 & 0 \\ 0 & 0 & 0 & 0 & 0 & 1 \\ 0 & 0 & \varepsilon & 0 & 0 & 0 \end{pmatrix} \qquad (5)$$

The graph of matrix $H_b$ is shown in FIG. 1, where the critical LSD that maximizes $\beta$ is explicitly indicated. As observed in the FIG. 1, $\varepsilon_{2,1}$ and $\varepsilon_{6,3}$ constitute the perturbation edges of the LSD under the maximum $\beta$ condition. Hence, $\mu = 2, \beta = 6, c(L) = 2, b_{2,6} = \varepsilon \times 1 \times 1 \times 1 \times \varepsilon = \varepsilon^2 \neq 0$. The number of remaining EPs of $H_b$ is 0.

$$H_c = \begin{pmatrix} 0 & 1 & 0 & 0 & 0 & 0 \\ 0 & 0 & 1 & 0 & 0 & 0 \\ \varepsilon & 0 & 0 & 1 & 0 & 0 \\ 0 & 0 & 0 & 0 & 1 & 0 \\ 0 & 0 & \varepsilon & 0 & \varepsilon & 1 \\ 0 & 0 & 0 & 0 & 0 & 0 \end{pmatrix} \qquad (6)$$

Matrix $H_c$ from Equation(6), it can be readily observed that the maximum number of vertices satisfying the LSD condition for this matrix is 4 in FIG. 1. $\varepsilon_{3,1}$ and $\varepsilon_{5,5}$ constitute the perturbation edges of the LSD. Hence, $\mu = 2, \beta = 4, c(L) = 2, b_{2,4} = \varepsilon \times 1 \times 1 \times \varepsilon = \varepsilon^2 \neq 0$. The number of remaining EPs of $H_c$ is 2.

**IV. Determine the splitting order of remaining EPs**

Having established a method to determine the number of remaining EPs, we now address a more subtle question: what is the splitting order of these remaining EPs when they are subjected to a further perturbation? The splitting behavior of the remaining EPs under a further perturbation can be decoded through an analysis of the complex eigenvalue trajectories.

**A. Theoretical Framework: Topological Analysis of Eigenvalue Trajectories**

This analysis begins by treating the perturbation strength as a complex variable, $\varepsilon = \alpha e^{i\theta}$, where $\alpha$ is a fixed, small positive constant, and the phase $\theta$ varies from 0 to $2\pi$. This parameterization renders the eigenvalues functions of $\theta$. As $\theta$ varies from 0 to $2\pi$, the eigenvalues evolve along continuous trajectories in the three-dimensional space spanned by $Re(\lambda)$, $Im(\lambda)$ and $\theta$. Subsequent analysis proceeds by examining the top-down views of these trajectories to make determinations.

(i) **Selection of Perturbation Magnitude $\alpha$**: An appropriate value of $\alpha$ must be chosen. Theoretically, a smaller $\alpha$ is preferable to adhere to the perturbation condition. However, an excessively small value makes visual details in the plot difficult to discern in the top-down views. Conversely, a value that is too large may cause the external and internal eigenvalue curves to intersect or overlap in the top views, violating the requirement for sufficiently small perturbations. A suitable $\alpha$ avoids both these issues.

(ii) **Identification of Exchange Types:** In the top view, we classify the connectivity of the eigenvalue trajectories: **Self-Exchange:** The trajectory of a single eigenvalue forms a closed loop, with its endpoint coinciding with its starting point. **Mutual-Exchange:** The terminus of one eigenvalue trajectory connects to the origin of another. A set of $v$ eigenvalues linked in this cyclic manner forms a closed loop.

(iii) **Quantification of the Winding Number ($d$):** The splitting order is governed by the winding number $d$, which counts how many times a trajectory winds around the base point(0,0) in the complex plane. For a closed curve $\gamma$ in the top view, the winding

number is formally defined as:
$$d = \frac{1}{2\pi i}\oint_\gamma \frac{dz}{z} \tag{7}$$
Where $\gamma$ is a closed curve, and $z$ is the complex plane equation of the closed curve $\gamma$ in the complex plane. While determining the exact analytic equation($z$) in the projection of the complex plane(top view) is undoubtedly challenging, this does not preclude the calculation of the winding number $d$. A more intuitive and practical method to determine the value of $d$ is to conceptually project a ray outward from the base point(0,0) and count how many times it intersects the closed trajectory $\gamma$. The number of intersections corresponds precisely to the value of $d$.

(iv)**Determination of the Splitting Order:** If the number of sets of eigenvalues constituting "exchange" is $v$, and the final number of winding forming the projection is $d$, then the eigenvalue splitting will follow $\lambda \sim \varepsilon^{d/v}$.

**B. Application to Example Matrices**

We take the concrete matrix $A, B, C, D$ as examples to illustrate the theory through graphical representation.

For Equation (8), since $A_p$ is subjected to only a single perturbation, specifically $\varepsilon_{5,1}$, and based on the conclusions in **Section III** regarding single perturbations, it can be deduced that the number of remaining EPs is 3. To further split these three remaining EPs, additional perturbations need to be introduced, while ensuring that no remaining EPs remain—meaning the system must attain full rank. Therefore, the applied perturbation must be constrained such that the LSDs of matrix contains a maximum of eight vertices, which is a necessary condition for eliminating all remaining EPs and thus achieving full rank. The perturbations $\varepsilon_{2,1}, \varepsilon_{4,3}$ and $\varepsilon_{8,5}$ form LSDs of $A$ encompassing all vertices as a consequence of $b_{3,8} = -\varepsilon^3$ in FIG. 4, which results in zero remaining EPs. To further investigate how the remaining EPs split, one can observe the top view of A in FIG. 2(b). From the method for determining the winding number described above, it can be seen that the black dashed curve starts from the base point (0,0) and subsequently intersects the inner curve as shown in FIG. 2(b) exactly twice. The mutual-exchange among the three eigenvalues leads to a splitting of the remaining EPs characterized by the scaling $\lambda_A \sim \varepsilon^{2/3}$.

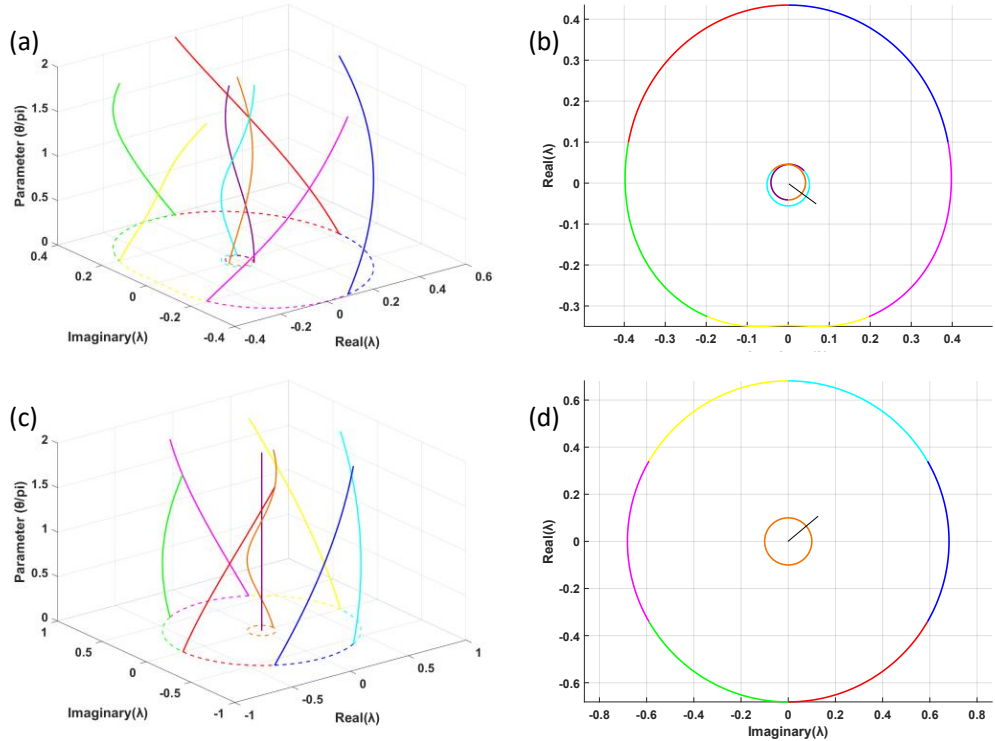

FIG. 2 3D views and top views of A and B, distinct colors represent distinct eigenvalue. where $\alpha$ of A is 0.01, $\alpha$ of B is 0.1 (a). Three-dimensional representation of the variation in the eigenvalues of matrix A, (b) Top view of the eigenvalue variation plot for matrix A. (c). Three-dimensional representation of the variation in the eigenvalues of matrix B, (d) Top view of the eigenvalue variation plot for matrix B.

The perturbation introduced in the evolution of Equation(8) aims to mutually exchange and fully split all remaining EPs. It should be noted that remaining EPs can also undergo partial splitting, which will be demonstrated through the evolution process described in Equation(9). The individual perturbation for $B_p$ is $\varepsilon_{6,1}$, hence the number of remaining EPs is 2. The perturbation $\varepsilon_{8,8}$ to be introduced is intended to cause incomplete splitting of the remaining EPs. As can be seen in the top view of diagram B in FIG. 2(d), the orange eigenvalues undergo self-exchange, thus resulting in a characteristic value splitting $\lambda_B \sim \varepsilon$. At this point, the number of remaining EPs for matrix B is reduced to one, as observed from the LSD of B in FIG. 4, where $b_{2,7} = \varepsilon^2$. A perturbation strategy involving the parameters $\varepsilon_{2,1}, \varepsilon_{4,3}$ and $\varepsilon_{8,5}$ applied to $D_p = B_p$ ensures the complete splitting of all remaining EPs in Eq.(11). The resultant configuration, which maximizes the number of vertex LSDs(FIG. 5), satisfies $b_{3,8} = -\varepsilon^3$. As can be seen in FIG. 3(d), the top view of D shows that the distributions of the orange and purple eigenvalues form a self-exchange, thus both satisfying $\lambda_D \sim \varepsilon$.

$$A_p = \begin{pmatrix} 0 & 1 & 0 & 0 & 0 & 0 & 0 & 0 \\ 0 & 0 & 1 & 0 & 0 & 0 & 0 & 0 \\ 0 & 0 & 0 & 1 & 0 & 0 & 0 & 0 \\ 0 & 0 & 0 & 0 & 1 & 0 & 0 & 0 \\ \varepsilon & 0 & 0 & 0 & 0 & 1 & 0 & 0 \\ 0 & 0 & 0 & 0 & 0 & 0 & 1 & 0 \\ 0 & 0 & 0 & 0 & 0 & 0 & 0 & 1 \\ 0 & 0 & 0 & 0 & 0 & 0 & 0 & 0 \end{pmatrix} \xrightarrow{+\varepsilon_{2,1}+\varepsilon_{4,3}+\varepsilon_{8,5}} A = \begin{pmatrix} 0 & 1 & 0 & 0 & 0 & 0 & 0 & 0 \\ \varepsilon & 0 & 1 & 0 & 0 & 0 & 0 & 0 \\ 0 & 0 & 0 & 1 & 0 & 0 & 0 & 0 \\ 0 & 0 & \varepsilon & 0 & 1 & 0 & 0 & 0 \\ \varepsilon & 0 & 0 & 0 & 0 & 1 & 0 & 0 \\ 0 & 0 & 0 & 0 & 0 & 0 & 1 & 0 \\ 0 & 0 & 0 & 0 & 0 & 0 & 0 & 1 \\ 0 & 0 & 0 & 0 & \varepsilon & 0 & 0 & 0 \end{pmatrix} \quad (8)$$

$$B_p = \begin{pmatrix} 0 & 1 & 0 & 0 & 0 & 0 & 0 & 0 \\ 0 & 0 & 1 & 0 & 0 & 0 & 0 & 0 \\ 0 & 0 & 0 & 1 & 0 & 0 & 0 & 0 \\ 0 & 0 & 0 & 0 & 1 & 0 & 0 & 0 \\ 0 & 0 & 0 & 0 & 0 & 1 & 0 & 0 \\ \varepsilon & 0 & 0 & 0 & 0 & 0 & 1 & 0 \\ 0 & 0 & 0 & 0 & 0 & 0 & 0 & 1 \\ 0 & 0 & 0 & 0 & 0 & 0 & 0 & 0 \end{pmatrix} \xrightarrow{+\varepsilon_{8,8}} B = \begin{pmatrix} 0 & 1 & 0 & 0 & 0 & 0 & 0 & 0 \\ 0 & 0 & 1 & 0 & 0 & 0 & 0 & 0 \\ 0 & 0 & 0 & 1 & 0 & 0 & 0 & 0 \\ 0 & 0 & 0 & 0 & 1 & 0 & 0 & 0 \\ 0 & 0 & 0 & 0 & 0 & 1 & 0 & 0 \\ \varepsilon & 0 & 0 & 0 & 0 & 0 & 1 & 0 \\ 0 & 0 & 0 & 0 & 0 & 0 & 0 & 1 \\ 0 & 0 & 0 & 0 & 0 & 0 & 0 & \varepsilon \end{pmatrix} \quad (9)$$

$$C_p = \begin{pmatrix} 0 & 1 & 0 & 0 & 0 & 0 & 0 & 0 \\ 0 & 0 & 1 & 0 & 0 & 0 & 0 & 0 \\ 0 & 0 & 0 & 1 & 0 & 0 & 0 & 0 \\ 0 & 0 & 0 & 0 & 1 & 0 & 0 & 0 \\ 0 & 0 & 0 & 0 & 0 & 1 & 0 & 0 \\ 0 & 0 & 0 & 0 & 0 & 0 & 1 & 0 \\ \varepsilon & 0 & 0 & 0 & 0 & 0 & 0 & 1 \\ 0 & 0 & 0 & 0 & 0 & 0 & 0 & 0 \end{pmatrix} \xrightarrow{+\varepsilon_{1,1}+\varepsilon_{2,2}+\varepsilon_{3,3}+\varepsilon_{4,4}+\varepsilon_{8,5}} C = \begin{pmatrix} \varepsilon & 1 & 0 & 0 & 0 & 0 & 0 & 0 \\ 0 & \varepsilon & 1 & 0 & 0 & 0 & 0 & 0 \\ 0 & 0 & \varepsilon & 1 & 0 & 0 & 0 & 0 \\ 0 & 0 & 0 & \varepsilon & 1 & 0 & 0 & 0 \\ 0 & 0 & 0 & 0 & 0 & 1 & 0 & 0 \\ 0 & 0 & 0 & 0 & 0 & 0 & 1 & 0 \\ \varepsilon & 0 & 0 & 0 & 0 & 0 & 0 & 1 \\ 0 & 0 & 0 & 0 & \varepsilon & 0 & 0 & 0 \end{pmatrix} \quad (10)$$

$$D_p = \begin{pmatrix} 0 & 1 & 0 & 0 & 0 & 0 & 0 & 0 \\ 0 & 0 & 1 & 0 & 0 & 0 & 0 & 0 \\ 0 & 0 & 0 & 1 & 0 & 0 & 0 & 0 \\ 0 & 0 & 0 & 0 & 1 & 0 & 0 & 0 \\ 0 & 0 & 0 & 0 & 0 & 1 & 0 & 0 \\ \varepsilon & 0 & 0 & 0 & 0 & 0 & 1 & 0 \\ 0 & 0 & 0 & 0 & 0 & 0 & 0 & 1 \\ 0 & 0 & 0 & 0 & 0 & 0 & 0 & 0 \end{pmatrix} \xrightarrow{+\varepsilon_{2,1}+\varepsilon_{4,3}+\varepsilon_{8,5}} D = \begin{pmatrix} 0 & 1 & 0 & 0 & 0 & 0 & 0 & 0 \\ \varepsilon & 0 & 1 & 0 & 0 & 0 & 0 & 0 \\ 0 & 0 & 0 & 1 & 0 & 0 & 0 & 0 \\ 0 & 0 & \varepsilon & 0 & 1 & 0 & 0 & 0 \\ 0 & 0 & 0 & 0 & 0 & 1 & 0 & 0 \\ \varepsilon & 0 & 0 & 0 & 0 & 0 & 1 & 0 \\ 0 & 0 & 0 & 0 & 0 & 0 & 0 & 1 \\ 0 & 0 & 0 & 0 & \varepsilon & 0 & 0 & 0 \end{pmatrix} \quad (11)$$

Self-exchange is not limited to a single winding; it can also involve multiple windings that alter the winding number. Through the evolution process described in Equation(10), $C_p$ retains only one remaining EPs. The perturbation term to be added ensures complete splitting, as can be observed from the LSDs of C, where $b_{5,8} = -\varepsilon^5$. As determined by the winding number method outlined earlier, the figure shows that the

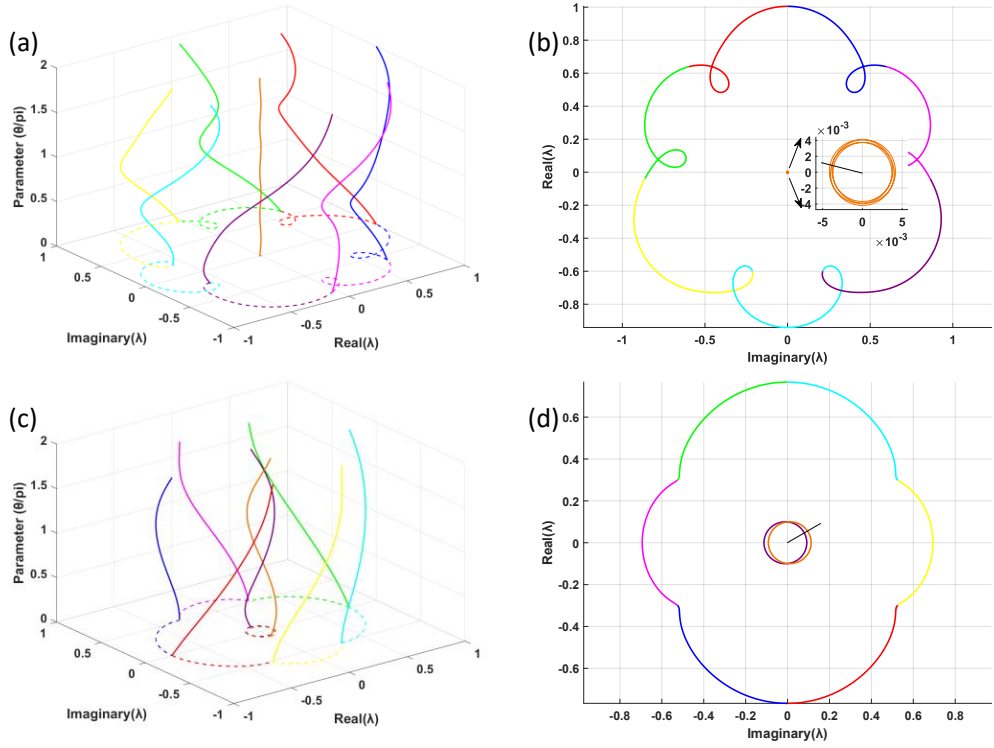

FIG. 3 3D views and top views of C and D, distinct colors represent distinct eigenvalue. where $\alpha$ of C is 0.25, $\alpha$ of D is 0.1 (a).Three-dimensional representation of the variation in the eigenvalues of matrix C,(b) Top view of the eigenvalue variation plot for matrix C (c).Three-dimensional representation of the variation in the eigenvalues of matrix D (d) Top view of the eigenvalue variation plot for matrix D.

spurious black branch crosses the orange eigenvalue trajectory on four occasions in FIG. 3(b). The eigenvalue splitting at the remaining EPs therefore scales as $\lambda_C \sim \varepsilon^4$. The four matrices are inductively summarized in the following schematic table:

| Initial State→Final State | Numbers of RemainingEPs of Initial State | Applied Perturbations | Winding number | Exchange Type |
|---|---|---|---|---|
| $A_p \rightarrow A$ | 3 | $\varepsilon_{2,1}, \varepsilon_{4,3}, \varepsilon_{8,5}$ | 2 | Mutual-Exchange |
| $B_p \to B$ | 2 | $\varepsilon_{8,8}$ | 1 | *Self-Exchange* |
| $C_p \to C$ | 1 | $\varepsilon_{1,1}, \varepsilon_{2,2}, \varepsilon_{3,3}, \varepsilon_{4,4}, \varepsilon_{8,5}$ | 4 | *Self-Exchange* |
| $D_p \to D$ | 2 | $\varepsilon_{2,1}, \varepsilon_{4,3}, \varepsilon_{8,5}$ | 1 | *Self-Exchange* |

## C. Splitting Characteristics under Real Perturbations

Although the preceding analysis models the perturbation as a complex quantity of the form $\varepsilon = \alpha e^{i\theta}$, it is a real-valued parameter without a significant phase component in some practical sensing scenarios. The method for determining the power-law scaling of eigenvalue splitting under real perturbations is consistent with the aforementioned method. The order of eigenvalue splitting does not depend on whether the perturbation is complex or real; however, when the perturbation is real-valued, it is more straightforward to determine whether the splitting occurs in the real parts, the imaginary parts, or both of the eigenvalues. The condition where $\theta = 0$ provides an ideal solution for this scenario: the projection method onto the complex plane can also determine whether the splitting occurs specifically in the real part, the imaginary part, or both of the all. The core principle is: The bifurcation behavior of an eigenvalue locus under parameter variation is dictated by its initial location. An eigenvalue initially located on the real axis undergoes a bifurcation in its real component, whereas one on the imaginary axis undergoes a bifurcation in its imaginary component. An eigenvalue starting at a generic point in the complex plane, away from the axes, experiences a concurrent bifurcation in both components.

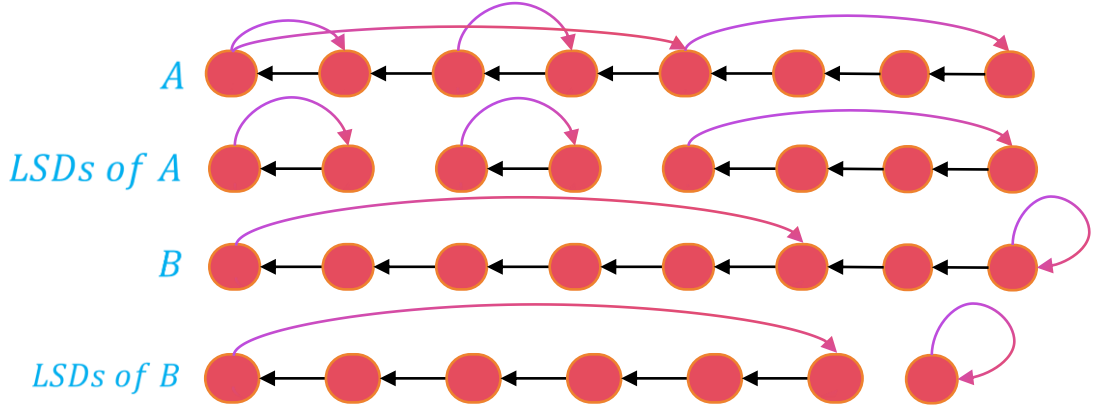

FIG. 4 The adjacency graphs of matrix A and matrix B respectively, along with their corresponding LSD diagrams that satisfy the maximum number of vertices. Red represents perturbation terms, and black represents the constant term 1.

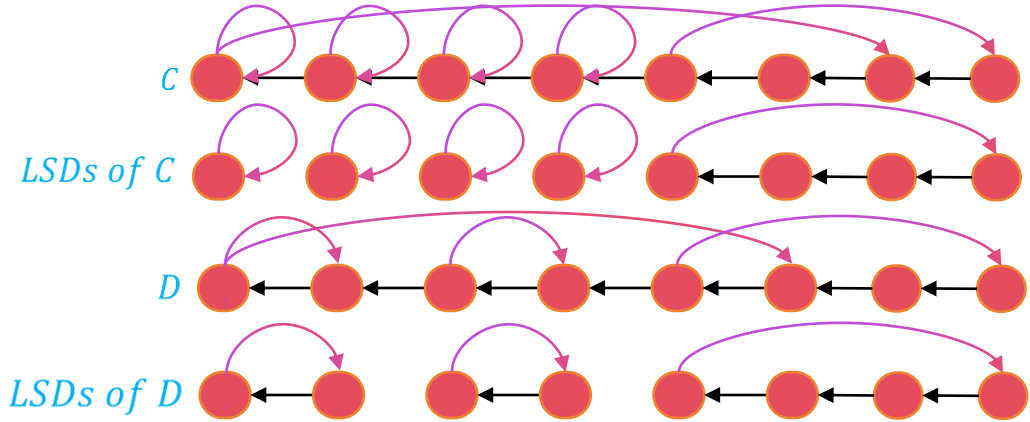

FIG. 5 The adjacency graphs of matrix C and matrix D respectively, along with their corresponding LSD diagrams that satisfy the maximum number of vertices. Red represents perturbation terms, and black represents the constant term 1.

A concrete demonstration of this principle will be provided using the previously introduced Eq.(8). For $B_p \to B$, Splitting of the remaining EPs is $\lambda_B \sim c\varepsilon$, according to the theory just discussed and FIG. 2(c) in top-view, the constant term $c$ must be purely real, since its starting point lies on the axis where the imaginary part is zero. This precisely indicates that the eigenvalues of the remaining EPs of B are characterized solely by a split in the real part.

Identifying whether the splitting occurs in the real or imaginary part of the eigenvalues provides key insight for elucidating the underlying physics of mode splitting[27] in sensing applications. In experiments of EP sensing in microcavities, Eq.(12) is commonly employed to quantify.

$$Q_{sp} = \frac{\text{Re}(\omega_{EP,2}) - \text{Re}(\omega_{EP,1})}{-\text{Im}(\omega_{EP,2}) - \text{Im}(\omega_{EP,1})} \quad (12)$$

If $Q_{sp} < 1$, it is experimentally challenging to observe mode splitting. Conversely, if $Q_{sp} > 1$, the splitting can be clearly observed. Understanding these concepts precisely explains why mode splitting is often accompanied by mode broadening. A comparative analysis with the Newton Polygon method is provided in **Appendix B**. The results confirm that our top-view approach not only offers a universal criterion for predicting bifurcations but also overcomes a critical limitation of the Newton Polygon method by enabling the determination of the bifurcation locus(real or imaginary), thereby providing a more complete characterization.

**V.Applicati ons of remaining EPs**

Remaining EPs provide a powerful framework for understanding anisotropic sensing behavior and the formation of Dirac exceptional points—a special class of EPs that do not traverse between PT-symmetric and PT-broken phases.As an illustrative example, consider the B-matrix subjected to two independent perturbations, denoted as $x$ and $y$, which induce anisotropic behavior in the system, where $x$ replaces the element $\varepsilon_{8,8}$ and $y$ replaces $\varepsilon_{6,1}$.Under the condition $x = 0$,perturbation solely along the $y$-direction manifests as a $\epsilon^{\frac{1}{6}}$ eigenvalue splitting.Conversely,when $y = 0$,perturbation solely along the $x$-direction manifests as a linear eigenvalue splitting.When perturbations $x$ and $y$ coexist simultaneously,the behavior along the $x$-direction directly coincides with the Dirac exceptional point discussed in the paper[28,29].By plotting the eigenvalues of Matrix B,we demonstrate to the reader the novel behavior of the Dirac exceptional point.In Fig5,It can be clearly seen that in the vicinity where $x$ approaches zero,the spectrum does not traverse the PT-symmetry and PT-symmetry broken phases.For the three-cavity direct-coupling model depicted in the figure on the right,the Hamiltonian is given by: $i\hbar \frac{\partial}{\partial t}\psi = H_0\psi$,Where $H_0 = \omega_0 I + \begin{pmatrix} ig & \kappa & 0 \\ \kappa & 0 & \kappa \\ 0 & \kappa & -ig \end{pmatrix}, g = \sqrt{2}\kappa, \kappa = 1$.The eigenvalues of $H_0$ is $\lambda_1 = \lambda_2 = \lambda_3 = \omega_0$.To facilitate subsequent analysis,we first subtract the eigenvalue multiplied by the identity matrix $I$ from the original matrix,then $\lambda_1 = \lambda_2 = \lambda_3 = 0$.Upon applying anisotropic perturbations $x$ and $y$,the Hamiltonian becomes

$$\begin{pmatrix} ig + x & \kappa & 0 \\ \kappa & 0 & \kappa \\ 0 & \kappa & -ig + y \end{pmatrix}$$

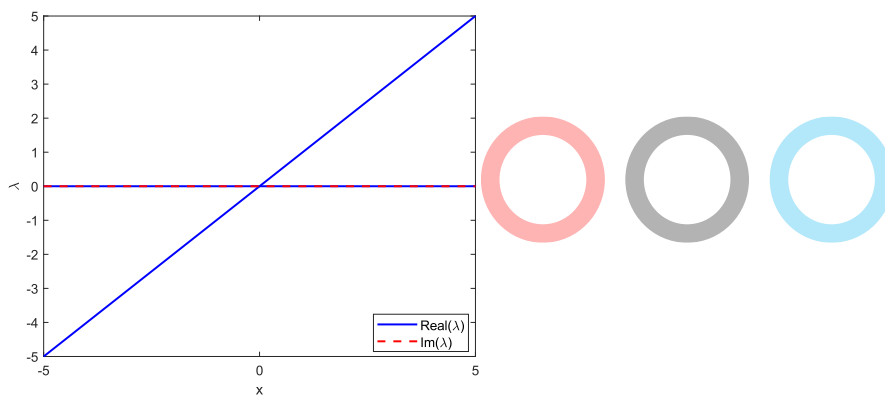

FIG. 6 Image of Dirac exceptional point eigenvalues(left),with the right figure showing a directly coupled model of three microcavities(red represents gain,gray no gain, blue indicates loss).

Based on our prior theoretical derivation,setting the determinant of this matrix to zero enables anisotropic behavior,yielding the condition $y = -x$.The analytical solutions for the eigenvalues can be expressed as:$\lambda_1 = -\sqrt{x(x + 2\sqrt{2}i)}, \lambda_2 = \sqrt{x(x + 2\sqrt{2}i)}, \lambda_3 = 0$.It is clearly observed that both $\lambda_1$ and $\lambda_2$ exhibit a $\epsilon^{\frac{1}{2}}$ splitting.All HOEPs can exhibit the characteristics of a Dirac EP,assuming that specific perturbations are applied,such as

$$\begin{pmatrix} ig & \kappa & 0 \\ \kappa & 0 & \kappa \\ -x & \kappa + ixg & x - ig \end{pmatrix}$$

The eigenvalues can be expressed as: $\lambda_1 = x, \lambda_2 = \lambda_3 = 0$. It can be observed that the formation of a Dirac exceptional point in HOEPs necessarily involves the presence of remaining EPs. The theoretical derivation for achieving this specific EP is straightforward. Simply transform the Hamiltonian $H_0$ into its Jordan canonical form via a similarity transformation. Apply a perturbation $x$ to an arbitrary diagonal entry of this canonical form matrix, then transform back via the inverse similarity transformation to obtain the new target matrix.

## VI. CONCLUSION

In summary, we rigorously demonstrated that if a n-order Hamiltonian containing EPn is perturbed with non-splitting eigenvalues, these points must form the remaining EPs. Furthermore, we proposed a method to determine the splitting order of perturbations via a top-down view in three-dimensional space, combined with a graph-theoretical LSDs approach to ascertain the number of remaining EPs. This approach effectively avoids the issues associated with numerical solutions in software for fitting splitting powers and also provides a deeper insight into the nature of eigenvalue splitting. We conducted cross-verification of the method proposed in this paper using Newton polygons.

## Appendix A

$n - 1 \leq rank(H) \leq n$, the proof is as follows:

**Upper Bound** ($rank(H) \leq n$): From the fundamental inequality of rank, $rank(H_0 + \varepsilon H_1) \leq rank(H_0) + rank(\varepsilon H_1)$, where the perturbation constant $\varepsilon (\varepsilon \neq 0)$ does not alter the rank of the matrix, i.e., $rank(\varepsilon H_1) = rank(H_1)$. Furthermore, as demonstrated in the paper[30], $H_0$ is Nilpotent Matrix, $H_0^{n-1} \neq 0, H_0^n = 0, rank(H_0) = n - 1$. Clearly, $H_1$ cannot be the zero matrix, as that would imply the system remains completely unperturbed, so $min(rank(H_1)) = 1$. As the rank cannot exceed the matrix dimension, it can be concluded that $rank(H) \leq rank(H_0) + rank(\varepsilon H_1) \leq n - 1 + 1 = n$.

**Lower Bound** ($n - 1 \leq rank(H)$): The proof of the lower bound leverages the fact that any matrix with an $n$ th-order EP can be transformed into its Jordan canonical form. Therefore, there exists $S^{-1} H_0 S = H^*$, where $H^*$ is a Jordan canonical form, $H^* = \begin{pmatrix} 0 & 1 & & 0 \\ & 0 & \ddots & \\ & & \ddots & 1 \\ 0 & & & 0 \end{pmatrix}$. It can be known that $H^*$ and $H_0$ are similar matrices, so $rank(H_0) = rank(H^*)$. Apply the following transformation to the equation: $S^{-1} HS = S^{-1} H_0 S + \varepsilon S^{-1} H_1 S$. Multiplying by an invertible matrix does not alter the rank, hence: $rank(H) = rank(S^{-1} HS)$. The key observation lies in the structure of $H^*$. The $(n - 1)$ th-order submatrix in the upper right corner of the $H^*$ matrix is denoted as $E = \begin{pmatrix} 1 & 0 & \cdots & 0 \\ 0 & 1 & \cdots & 0 \\ \vdots & \vdots & \ddots & \vdots \\ 0 & 0 & \cdots & 1 \end{pmatrix}$. We now consider the effect of the perturbation: Assuming $H^*$ is subjected to arbitrary perturbations, If the perturbation is only applied to the exterior of sub-matrix $E$, according to the definition of matrix rank (if there exists an $r$-th order sub-matrix whose determinant is not zero, and the determinant of any $(r + 1)$-th order sub-matrix is zero, then the rank of this matrix is $r$), it can be concluded that $n - 1 \leq rank(H)$. If the perturbation is not limited to the exterior of the $E$ matrix,but exists both internally and externally.For the sub-matrix,there is $E + \varepsilon H_2$,where $H_2$ is the upper-right $(n-1)$rd-order sub-matrix of $S^{-1}H_1S$ .In this case,the expression for the eigenvalue of sub-matrix is $1 + \lambda(\varepsilon)$.Regardless,the eigenvalue is certainly not zero,because as $\varepsilon \to 0$,there must be $\lambda(\varepsilon) \to 0$.Therefore,it can be shown that $rank(E + \varepsilon H_2) = n - 1$,which necessarily implies $n - 1 \leq rank(H)$.

## Appendix B

We can utilize Newton polygons to verify the validity of the criterion for eigenvalue splitting. The Newton polygon[31] is a powerful geometric device that constructs a convex polygonal chain from the valuations of the coefficients of a polynomial(or power series) relative to a given valuation. This structure encapsulates critical arithmetic information about the roots of the polynomial,particularly concerning their valuations. The steps to determine the ramification index using the Newton polygon are as follows: One must compute the algebraic expression of $p(\lambda, \epsilon) = \det(H - \lambda I)$. After computation, one should obtain a concrete algebraic expression for $p(\lambda, \epsilon)$,which can be represented in a standard form such as $\sum_{q,w} a_{qw} \lambda^q \epsilon^w$. Plot all existing ordered pairs $(q, w)$ from the algebraic expression on a two-dimensional plane.The smallest convex shape that contains all the points plotted is called the Newton polygon. After constructing the Newton polygon,we select edges such that all other vertices lie above or to the right of the line defined by each chosen edge. The negative value of the slope of this line segment then serves as the basis for determining the degree of splitting. The formula for calculating slope is

$$k_j = \frac{w_j - w_{j-1}}{q_j - q_{j-1}} \tag{13}$$

Sometimes, more than one edge satisfies the condition,which means we may obtain multiple slope values. Certain edges dominate the primary splitting, while the remaining edges govern the splitting of the remaining exceptional points. The expressions for $p(\lambda, \epsilon)$

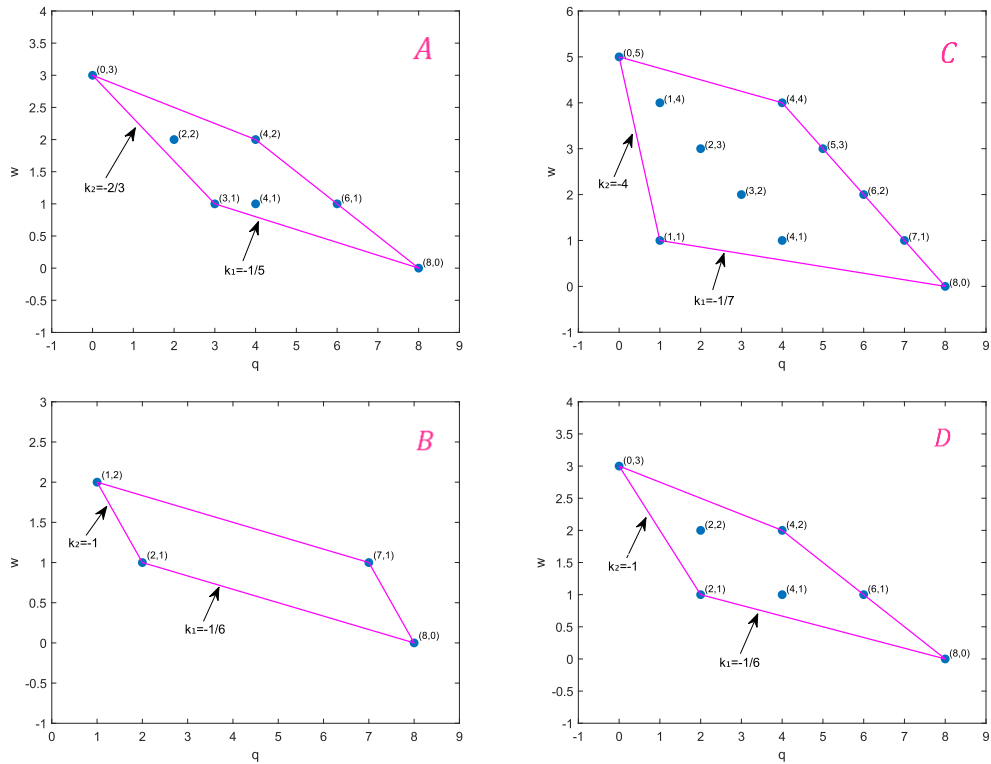

FIG. 7 Newton polygon images corresponding to matrices A, B, C,and D respectively.

of matrices $A, B, C$ and $D$ are calculated respectively as follows:
$$\det(A - \lambda I) = -\epsilon^3 + \epsilon^2\lambda^4 + 2\epsilon^2\lambda^2 - 2\epsilon\lambda^6 - \epsilon\lambda^4 - \epsilon\lambda^3 + \lambda^8 \quad (14a)$$
$$\det(B - \lambda I) = \epsilon^2\lambda - \epsilon\lambda^2 - \epsilon\lambda^7 + \lambda^8 \quad (14b)$$
$$\det(C - \lambda I) = -\epsilon^5 + \epsilon^4\lambda^4 + 4\epsilon^4\lambda - 4\epsilon^3\lambda^5$$
$$-6\epsilon^3\lambda^2 + 6\epsilon^2\lambda^6 + 4\epsilon^2\lambda^3 - 4\epsilon\lambda^7 - \epsilon\lambda^4 - \epsilon\lambda + \lambda^8 \quad (14c)$$
$$\det(D - \lambda I) = -\epsilon^3 + \epsilon^2\lambda^4 + 2\epsilon^2\lambda^2 - 2\epsilon\lambda^6 - \epsilon\lambda^4 - \epsilon\lambda^2 + \lambda^8 \quad (14d)$$

As shown in Figure 7, the slopes of the eligible edges in the Newton polygon of matrix A indicate the degree of eigenvalue splitting. Specifically, the dominant eigenvalues scale as $\lambda \sim \epsilon^{\frac{1}{5}}$, whereas after introducing perturbations to the remaining EPs in the evolution process described by equation(14), the eigenvalues satisfy $\lambda \sim \epsilon^{\frac{2}{3}}$. A similar principle applies to matrices B, C, and D as well. The results for the splitting degree obtained using Newton polygons show strong agreement with those derived from the winding number method described above, indicating that the top-view method is highly useful. While Newton polygons can effectively predict the exponents at which eigenvalues split, the last two figures(B and D) clearly show a limitation of this method. It is clearly visible from the figure that the Newton polygon does not directly reveal the nature of the splitting concerning their real and imaginary parts. This necessitates explicit algebraic computation to ascertain the outcomes, the steps of which are schematically shown below: First, calculate the relationship for the fitted curve based on the slopes of the Newton polygon. Identify the corresponding terms in the characteristic polynomial($p(\lambda, \epsilon)$) such that their sum is zero. Substitute the fitted equation into the resulting characteristic polynomial, solve for the possible values of c, and thus determine the number of eigenvalue branches in this case. Using the B and D matrices from the last two figures as examples.

$$\lambda \sim c\epsilon^{\frac{1}{6}} \quad (15a)$$
$$-\epsilon\lambda^2 + \lambda^8 = 0, -c^2\epsilon^{\frac{4}{3}} + c^8\epsilon^{\frac{4}{3}} = 0 \quad (15b)$$
$$\lambda \sim c\epsilon \quad (15c)$$
$$-\epsilon^3 - \epsilon\lambda^2 = 0, -\epsilon^3 - c^2\epsilon^3 = 0 \quad (15d)$$
$$\epsilon^2\lambda - \epsilon\lambda^2 = 0, c\epsilon^3 - \epsilon^3 c^2 = 0 \quad (15e)$$

Where Equation(15a) represents the split fitting for the dominant term, Equation(15c) represents the split fitting for the remaining exceptional point under perturbation, the left-hand side of Equation(15b) corresponds to extracting the relevant term from the characteristic polynomial, and the right-hand side of Equation(15b) is obtained by substituting Equation(15a). Equations(15d) and (15e) are derived following the same principle, and so on. From Equation(15b), solving the right-hand side yields two solutions: $c^2 = 0$(meaningless, thus discarded) or $c^6 = 1$. This implies the existence of six solution branches. Similarly, solving Equation(15d) gives $c^2 = -1$, etc. This implies that only the imaginary parts of the eigenvalues undergo splitting. Similarly, solving for the case of (15e) yields c=1, meaning one eigenvalue remains unsplit. It can be seen that the Newton polygon alone cannot directly reveal which the splitting of eigenvalues occurs in their real or imaginary parts. These aspects must be determined through rigorous computation. Our proposed Top view-based Theoretical Framework can effectively resolve such scenarios.